\documentclass[aps,prl,twocolumn,showpacs]{revtex4}

\usepackage{graphicx}
\usepackage{dcolumn}
\usepackage{bm}
\usepackage{subfigure}

\begin{document}

\preprint{APS/123-QED}

\title{Resonant Activation of a Current-Biased Josephson Junction Near the Classical-Quantum Crossover}

\author{Z. E. Thrailkill}
\author{J. G. Lambert}
\author{S. A. Carabello}
\author{R. C. Ramos}

\affiliation{%
Low Temperature Lab, Department of Physics, Drexel University, Philadelphia, PA 19104 USA
}%

\date{\today}

\begin{abstract}
We examine the resonant activation of a current-biased Josephson junction near the crossover temperature in order to show the way in which the device transitions from the classical regime to the quantum regime.  We use microwaves to probe the quantum energy states that exist in the potential well.  The quantum features are visible until the junction is heated up to the crossover temperature, at which point the line widths of the energy levels overlap and become indistinguishable from one another.  When well above this temperature, the junction behaves classically when resonantly activated with microwaves. 
\end{abstract}

\pacs{74.50.+r, 85.25.Cp, 03.65.-w, 05.30-d, 85.25.Am}
\maketitle

\section{\label{sec:level1}INTRODUCTION\protect\\}

The superconducting Josephson junction is a solid state device, consisting of two superconductors separated by a weak link.  Since its development, the junction and circuits containing it have played significant roles in both basic physics and applied technology \cite{devoret-2004}.  Its dynamics, as a function of the macroscopic phase difference across the junction, are described by a washboard potential.  When cooled to sufficiently low temperatures and electrically isolated, these devices exhibit quantized energy levels within its potential well.  The Josephson junction has already received a great deal of attention when operated at temperatures in the quantum regime.  They have been tested for quantum mechanical features such as energy level spectroscopy \cite{guozhu:104531,PhysRevB.71.064512}, Rabi oscillations \cite{PhysRevLett.95.067001,PhysRevLett.89.117901,1367-2630-10-7-073026}, time evolution of quantum states \cite{YangYu05032002,Pashkin-2003}, and quantum entanglement \cite{A.J.Berkley06062003,R.McDermott02252005,steffen:050502,PhysRevLett.94.027003,Pashkin-2003,PhysRevB.68.024510,PhysRevLett.94.090501,grajcar:047006}.  However, if the temperature is high enough, the junction will behave in a classical manner.  In this paper, we will investigate what happens when the junction is between these two sets of conditions.  

We will examine the crossover between the quantum and classical behavior of the Josephson junction \cite{PhysRevLett.93.107002,PhysRevLett.79.3046}.  To do this, we perform spectroscopic experiments by applying microwave signals to the junction, and observing the switching events to the finite voltage state.  Based on experimental results, we find that the system behaves in a way that exhibits both quantum and classical properties.  Specifically, we present data near the crossover temperature demonstrating the transition between quantum and classical behavior.

\section{\label{TH}Theory\protect\\}

\begin{figure}[htbp]
\subfigure[]{
\includegraphics[width=1.5in]{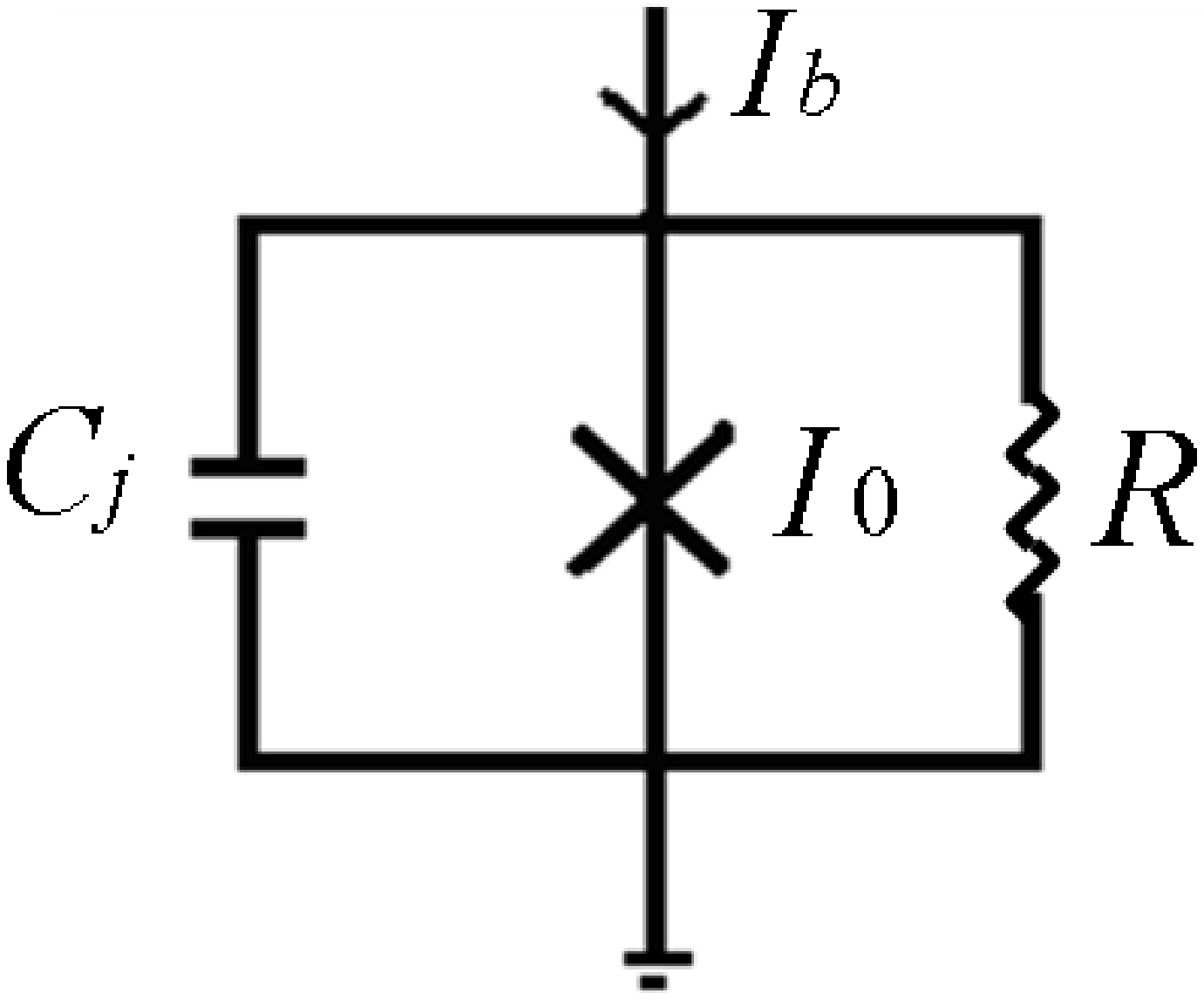}
\label{fig1a}}\\
\subfigure[]{
\includegraphics[width=1.5in]{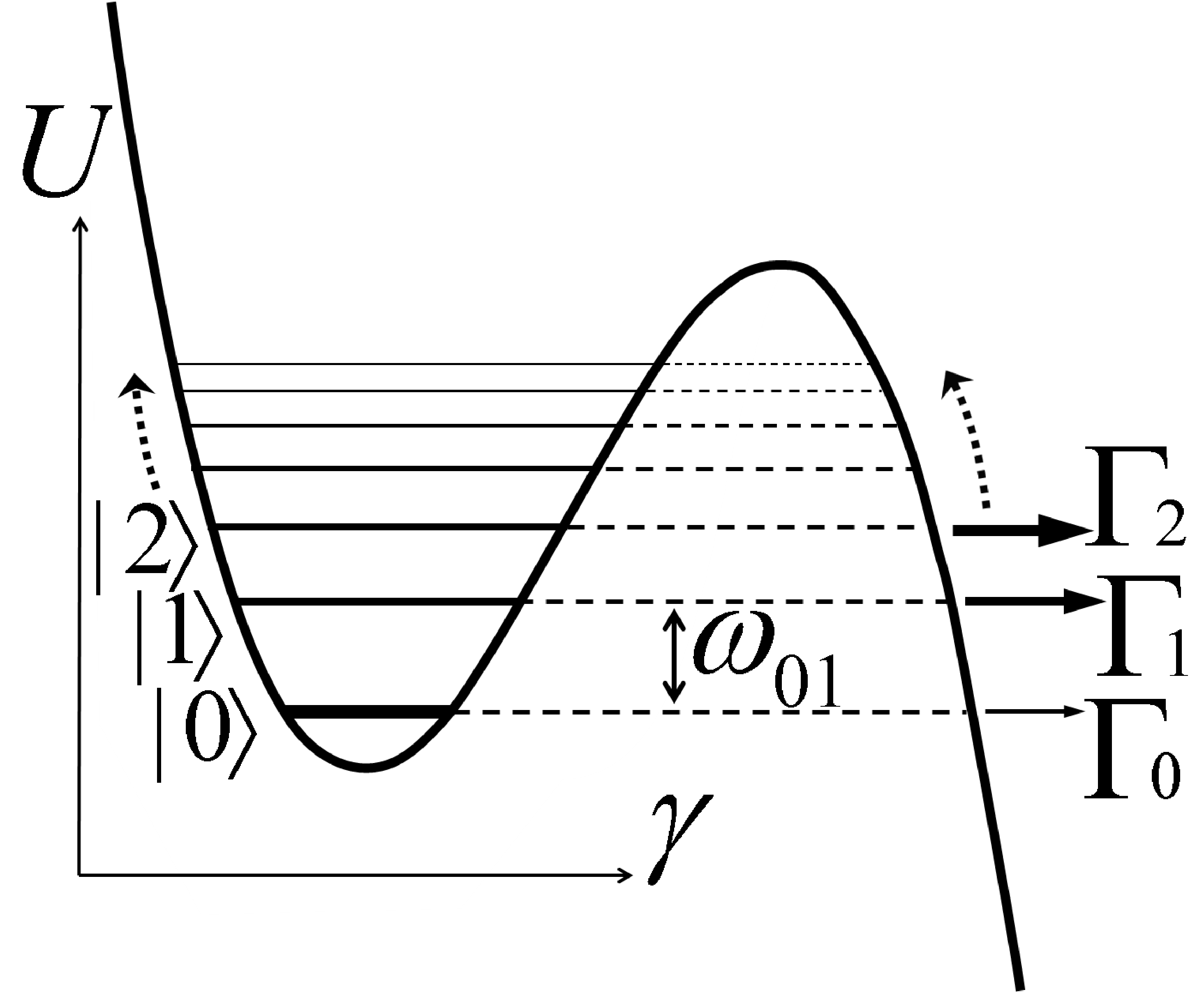}
\label{fig1b}}
\subfigure[]{
\includegraphics[width=1.5in]{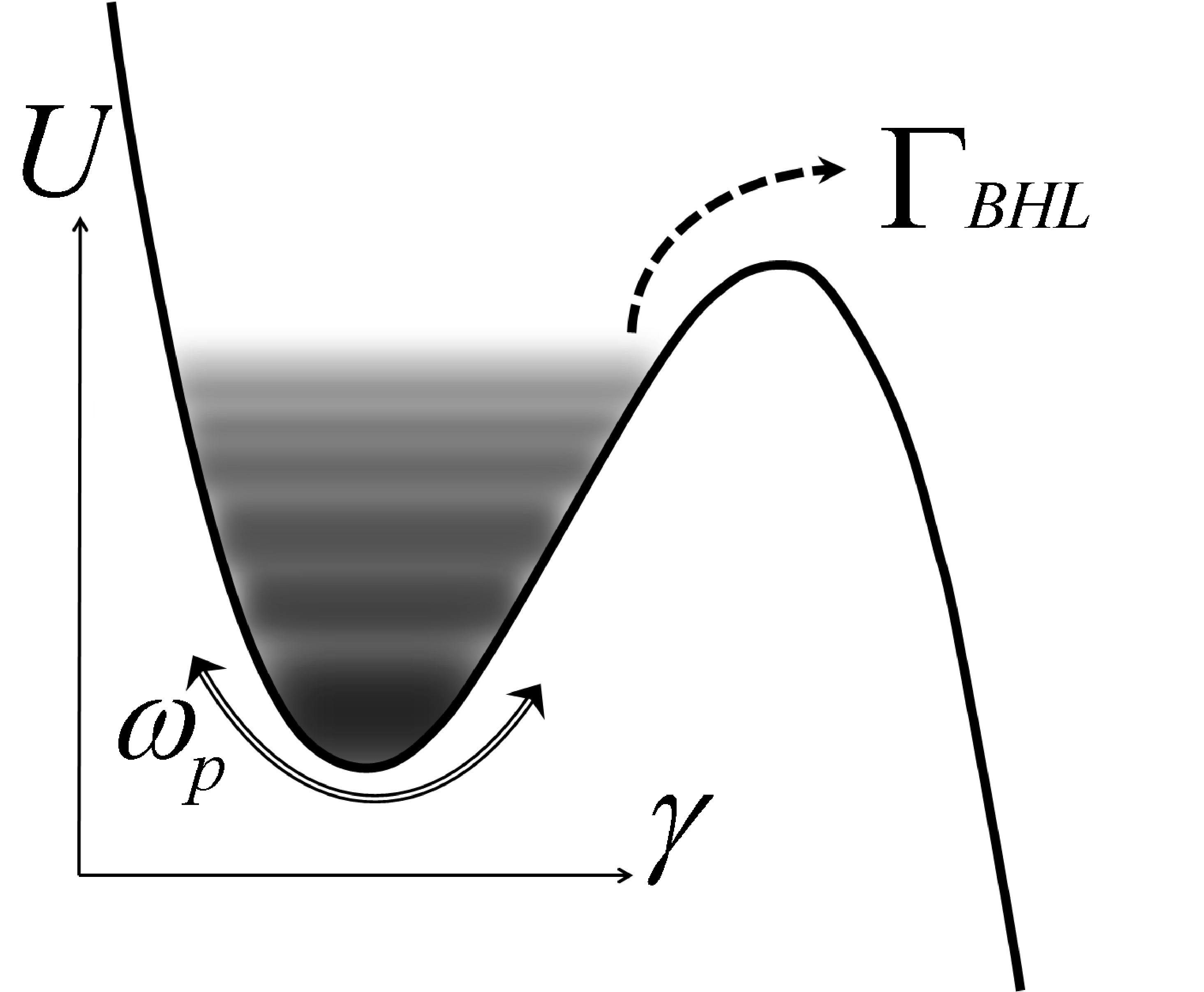}
\label{fig1c}}
\caption{ (a) The RCSJ model for a Josephson junction.  The junction is represented by the cross shunted by its capacitance and resistance.  (b) The washboard potential of the junction showing quantum states.  At sufficiently low temperatures only the lowest states are populated and thermal escape out of the well is minimized due to a lack of thermal energy.  In this quantum regime the primary escape mechanism is tunneling through the barrier.  Higher energy levels will escape at higher rates.  Therefore, if microwaves are applied at the resonance frequency, $\omega_{01}$, the first excited state will become populated, and an enhancement in the escape rate will be observed.  (c) At higher temperatures, more energy states become populated, causing the system to exist in a superposition of energy states, allowing it to behave as though there was a continuum of energy states.  Under resonant activation this system can behave classically.}
\end{figure}

The Josephson junction can be modeled by the RCSJ model, as represented by Fig.~\ref{fig1a}, where $R$ is the total parallel shunt resistance of the junction, $C_j$ is the junction capacitance, $I_b$ is the bias current through the junction, and $I_0$ is the critical current \cite{Xuthesis}.  The dynamics of the current-biased Josephson junction is governed by the washboard potential, $U(\gamma ) =  - \frac{{\Phi _0 }}{{2\pi }}(I_0 \cos (\gamma ) + I_b \gamma )$, as shown in Fig.~\ref{fig1b}, where  $\gamma$ is the phase difference across the junction, and $\Phi _0 =h/(2e)= 2.07x10^{-15}$ Wb is the flux quantum.  For $I_b < I_0$ the system's phase is trapped in the potential well, resulting in a zero voltage drop across the device.  

Analysis of the classical behavior of the device involves a 'phase particle' that is trapped in the well which oscillates with a plasma frequency~\cite{PhysRevB.9.4760} 

\begin{equation}
\label{eqwp}
\omega _p  = \sqrt {\frac{{2\pi I_0}}{{\Phi_0C_j}}} \left[ {1 - \left( {\frac{{I_b }}{{I_0 }}} \right)^2 } \right]^{\frac{1}{4}}.  
\end{equation}

\noindent In reality, the classical system is still comprised of quantum energy levels, but with enough thermal energy to smear the state population amongst several levels.  With a large number of superposition states, the junction will effectively have a continuum of energies and behave classically.  If the system is cooled down to very low temperatures, the range of energies that the junction can attain due to a superposition of thermally excited states, decreases, ultimately approaching a set of quantized energy level states.

From the quantum perspective, metastable energy states exist in the well.  The energy spacing between the ground state, $\left|0\right\rangle$, and the first excited state, $\left|1\right\rangle$, will depend on the barrier height.  The spacing may be approximated by~\cite{Berkleythesis}

\begin{equation}
\label{eqw01}
\omega_{01}\cong\omega _p\left(1-\frac{5\hbar\omega _p}{36\Delta U}\right).
\end{equation}

\section{\label{ex}Experiment\protect\\}

Spectroscopy was performed by linearly ramping the bias current through the critical current and detecting the switching event from the zero voltage state to the finite voltage state.  This procedure was repeated on the order of $10^{4}$ to $10^{5}$  times.  Histograms of the switching events were produced.  To prepare the junction in its ground state, it was cooled in a helium dilution refrigerator with a base temperature of 18mK.  The device was enclosed in an aluminum sample box which becomes superconducting below 1 K, shielding the junction from external magnetic fields.  It was further enclosed in a Cryoperm 10 cylinder from Amuneal Manufacturing Corp. to shield the sample from DC magnetic fields.  The current bias line of the junction was thermalized and heavily filtered down to a frequency of about 16MHz using 3 meters of Thermocoax wire, and home-made LC and copper powder filters.  At room temperature, low-noise amplifiers and a Schmitt trigger powered by batteries were used to detect switches to the finite voltage state.  These electronics were placed atop the cryostat of the dilution refrigerator, which was all enclosed in a copper fabric shroud to electrically shield ambient RF noise.  Detection electronics were isolated using a fiber optic cable that transmitted a TTL pulse for each switching event detected.  Microwaves were coupled via an antenna in the sample box, suspended above the junction.  A 30db fixed attenuator was used to thermalize the microwave line.  

\begin{figure}[htbp]
\includegraphics[width=3.3in]{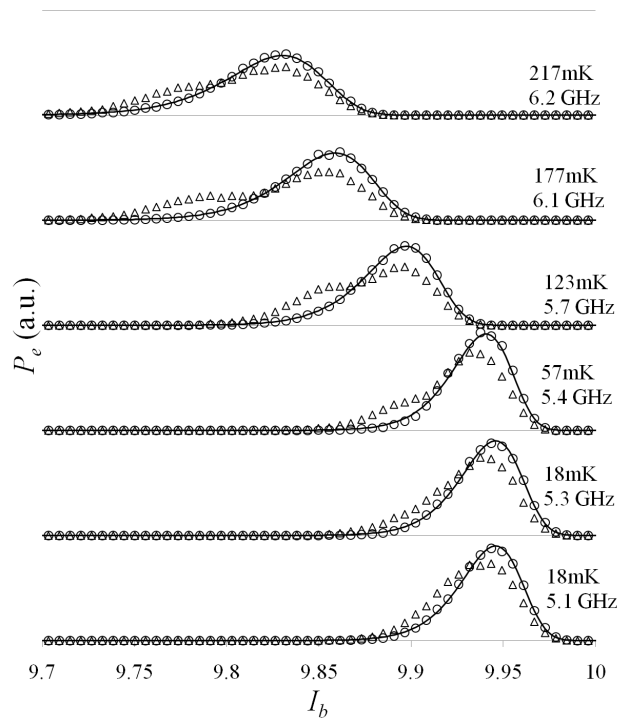}
\caption{\label{fig6} Histograms of the escape events as a function of bias current, offset for clarity.  The circles correspond to points without microwaves and the triangles are with microwaves.  Theoretical fits are shown through each of the RF off curves.  }
\end{figure}

Data was taken at various temperatures ranging from 18mK to 2.5K.  Some data taken at base and elevated temperatures can be seen in Fig.~\ref{fig6}.  The microwave frequency used to excite the junction was tuned as the temperature increased in order to maintain a visible resonance.  As the temperature increases, the histogram peaks shift to the left due to the increase in thermal noise that causes premature escapes from the washboard well.  

\section{\label{HT}High Temperature Results\protect\\}

In order to analyze the data, we must first find the parameters of our system~\cite{PhysRevB.35.4682}.  For our junctions, $C_j$ is about 5pf.  We use theoretical fits in order to find $R$, $I_0$, and $T$.  In the limit that the system is above the crossover temperature $T\equiv\hbar\omega_{p0}/2\pi k_b$, the total escape rate is governed by the equation~\cite{Kramers1940284} 

\begin{equation}
\label{bhl}
\Gamma_{BHL} = a_t\frac{{\omega p}}{{2\pi }}\exp \left( { - \frac{{\Delta U}}{{k_b T}}} \right),
\end{equation}

\noindent where $a_t  = 4/[\sqrt {1 + 5Qk_b T/9\Delta U}  + 1]^2  $ is a factor that takes into account population depletion for energies above the barrier height~\cite{PhysRevB.28.1268}, and $Q=\omega_p RC$ is the quality factor.  For each temperature, the escape rate $\Gamma_{BHL}$ is fitted to the data with no microwaves.  In order to find the best fit, the values for $R$, $I_0$, and $T$ in Eq. (\ref{bhl}) were iterated through a range of values, providing a set of theoretical data points, $P_{ti}$, that can then be compared to experimental data points represented by escape counts, $P_{ei}$.  The final theoretical points are plotted with the experimental data in Fig.~\ref{fig6}.  For each iteration, the quality of the fit was determined by the equation \cite{PhysRevB.37.1525}:

\begin{equation}
\label{xi}
\chi^{2}=\sum(P_{ei}-P_{ti})^{2}/\sigma_i^{2}
\end{equation}

\begin{figure}[htbp]
\includegraphics[width=3.3in]{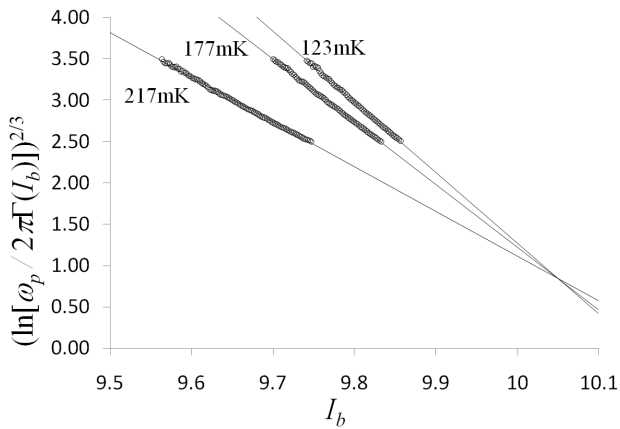}
\caption{\label{fig7} In order to find the critical current for the junction, $(ln[\omega_p/2\pi\Gamma_e])^{2/3}$ is plotted as a function of bias current.  The linear fits intersect at the theoretical critical current at about $I_0=10.06\mu A$.  This provides a way to constrain the critical current parameter used in the fits.}
\end{figure}

The best fit was determined by the parameters that minimized $\chi^{2}$.  However, there are many different combinations of $R$, $I_0$, and $T$ that give similar $\chi^{2}$ values.  It was helpful to have an estimate of at least one of the parameters in order to constrain the system.  The iterative technique above can be used to find an approximate range for $I_0$ that can be used in Eq. (\ref{eqwp}).  Using $\omega_p$ and the experimental escape rate $\Gamma_e$ we plot $(\ln[\omega_p/2\pi\Gamma_e])^{2/3}$ as a function of $I_b$ for the data at various temperatures \cite{PhysRevB.35.4682,PhysRevB.9.4760}.  As seen in Fig~\ref{fig7}, these plots are linear and when fitted with a least squares line, they intersect at one point.  The intersection gives us a more accurate value for $I_0$. We found the critical current to be $I_0=10.06$ $\mu A$.

Figure~\ref{fig6} shows the escape histogram data taken at different temperatures, and their fitted curves.  The fits predict $R=1670$ $\Omega$. This results in a quality factor of $Q\approx300$.  Using the critical current derived from the fit, we can find $\omega_p$ and $\omega_{01}$.  

\begin{figure}[htbp]
\includegraphics[width=3.3in]{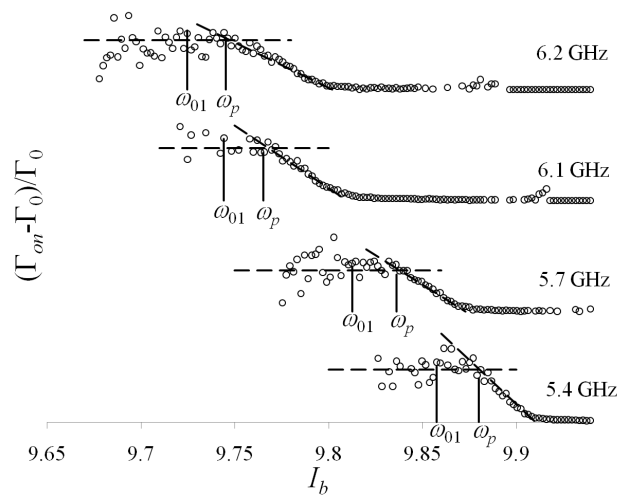}
\caption{\label{fig4} Enhancement plots corresponding to the data in Fig.~\ref{fig6}.  The location of the $\omega_{01}$ and $\omega_p$ resonances are shown for each plot.  The non-Lorentzian shape is indicative of classical behavior~\cite{rotoli:144501}.  The step-like structure is caused by the addition of energy via off resonance microwaves.  This is allowed due to the superposition of states caused by the higher temperatures, resulting in classical behavior.  Otherwise the microwaves would only cause an enhancement when they are resonant with one of the quantum energy level spacings.  With significantly large line widths, the enhancement would have the step-like structure, but the elbow would line up with the resonant energy spacing.  The elbow in the enhancements shown here clearly line up with $\omega_p$, indicating classical resonance. }
\end{figure}

The data can then be used to produce escape rate curves for both the RF off, $(\Gamma_0)$, and RF on, $(\Gamma_{on})$ data.  From these, an enhancement graph can be made by taking $(\Gamma_{on}-\Gamma_0)/\Gamma_0$.  If the system behaves classically, then the enhancement will not be a Lorentzian, as is the case in a quantum system.  Instead, it will have a step-like structure~\cite{rotoli:144501,PhysRevB.35.4682} as shown in Fig.~\ref{fig4}.  Here, the enhancements for the elevated temperatures from Fig.~\ref{fig6} are plotted.  The fitted temperatures are 106mK, 160mK, 212mK, and 252mK, from bottom to top.  These elevated temperatures indicate that there is heating at the junction or excess current noise.  

The step-like structure comes from the energy levels above $\left|1\right\rangle$ having enough thermal population to allow the state of the junction to be described by a superposition of the quantum states.  This allows for classical motion of the phase 'particle' trapped in the well.  Evidence of this superposition of states is seen in Fig.~\ref{fig4}, where the location of the predicted $\omega_{01}$ peak is beyond the elbow of the plateau in the enhancement, while $\omega_p$ is at the elbow.  This indicates that the system is in the primarily classical regime, which is expected at these elevated temperatures.   

\section{\label{NC}Quantum/Classical Crossover\protect\\}

As the temperature of the device is lowered to the crossover temperature, quantum features begin to appear.  In this regime, escape rates from individual quantum energy levels become comparable to the thermal escape rate.  To better understand what is happening, consider the junction in its quantum ground state.  Here, quantum tunneling dominates the escapes and the enhancement will be a single Lorentzian peak.  As the temperature increases, thermal excitations will populate the first excited state enough to allow a second peak corresponding to the $1\rightarrow2$ transition.  As the temperature is further increased, more peaks will appear, and the switching current peak will broaden.  Eventually, these broad peaks will overlap and form the step-like structure that is seen in the case of the classical system.  However, the elbow of the enhancement should still be at the $\omega_{01}$ peak \cite{1211764}.  It is not until the temperature increases further that the system will have enough thermal population in the excited states to form the superposition necessary to allow for classical motion of the phase.  

\begin{figure}[htbp]
\includegraphics[width=3.3in]{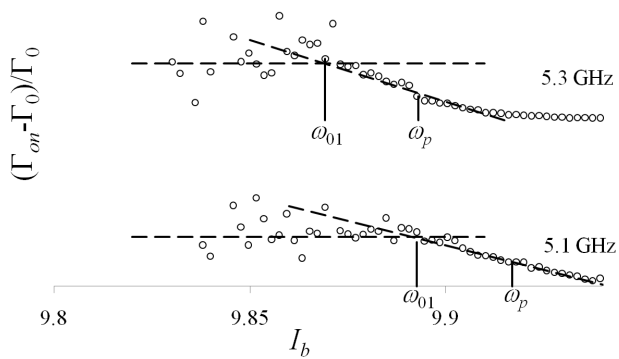}
\caption{\label{fig5} Enhancement plots for the data at base temperature in Fig.~\ref{fig6}.  As in Fig.~\ref{fig4}, the bias currents corresponding to a resonance with $\omega_{01}$ and $\omega_p$ are shown.  At lower temperatures, the system approaches the quantum crossover, though there is still the step-like structure as before.  The key difference is that the elbow of the plot corresponds to the $\omega_{01}$ resonance.  Therefore, the step structure is caused by broadened quantum energy levels rather than classical motion.  The enhancement peaks for the various excited states are wide enough at these temperatures that they all merge together to form a step.  The enhancement then drops off after the junction is biased past the resonance with the final energy state.  This indicates that the system is operating near the classical/quantum crossover point.}  
\end{figure}

Fig.~\ref{fig5} shows some of our base temperature data taken from the same junction as above.  The data was fit to an effective temperature of 101mK, which indicates that the heating, or current noise, has saturated at this point.  The crossover temperature is about 100mK, so even with the elevated temperature we are still close enough to the crossover point and expect to see an indication of quantum behavior.  The data in Fig.~\ref{fig5} shows the expected step-like features.  Clearly, the elbows of the steps are now aligned with the $\omega_{01}$ peaks, evidence that we are approaching the quantum regime.  This indicates that the system dynamics are being governed by overlapping quantum line widths rather than a superposition of energy states.  

In conclusion, we have presented experimental data that shows evidence of the crossover from classical to quantum behavior.  The first distinctive change that occurs when making the transition is the observed shift of the $\omega_p$ and $\omega_{01}$ resonance points.  In the high temperature regime, the escape enhancement is governed by the classical plasma frequency.  As the temperature is decreased, approaching the crossover point, the enhancement will become governed by the resonances corresponding to the quantum energy level spacings.  This provides us with a better understanding of the intermediate steps between quantum and classical behavior.  In the future, we hope to provide a more detailed look into the interplay of classical and quantum dynamics in Josephson junctions.

\bibliography{apssamp}

\end{document}